\documentclass[10pt,journal,compsoc]{IEEEtran}
\ifCLASSOPTIONcompsoc
\usepackage[nocompress]{cite}
\else
\usepackage{cite}
\fi

\usepackage{amsmath,amssymb}
\usepackage{multirow}
\usepackage{xcolor}
\usepackage{graphicx,algorithm,algcompatible,textcomp,tabularx}
\usepackage{enumitem} 
\usepackage[font={small}]{caption,subcaption}

\newcommand{\revtr}[1]{{\color{black}#1}}

\def\BibTeX{{\rm B\kern-.05em{\sc i\kern-.025em b}\kern-.08em
		T\kern-.1667em\lower.7ex\hbox{E}\kern-.125emX}}
\makeatletter
\renewcommand\paragraph{\@startsection{paragraph}{4}{\z@}%
	{1.5ex plus .2ex minus .3ex}%
	{-0em}%
	{\normalsize\bf}}
\makeatother

\usepackage{url}

\begin{document}
	
	\title{Invariant Feature Learning for Sensor-based Human Activity Recognition}
	
	\author{Yujiao~Hao,
		Rong~Zheng,~\IEEEmembership{Senior~Member,~IEEE}
		and~Boyu~Wang
		\IEEEcompsocitemizethanks{\IEEEcompsocthanksitem Yujiao Hao and Rong Zheng are with the Department of Computing and Software, McMaster University, Hamilton, Canada. E-mail: haoy21@mcmaster.ca; rzheng@mcmaster.ca.\protect\\
			\IEEEcompsocthanksitem Boyu Wang is with Department of Computer Science, Western University, London, Canada. E-mail: bwang@csd.uwo.ca.}
	}
	
	\markboth{IEEE TRANSACTIONS ON MOBILE COMPUTING, VOL. xx, NO. x, xxx 2020}%
	{Shell \MakeLowercase{\textit{Hao et al.}}: Invariant Feature Learning for Sensor-based Human Activity Recognition}
	
	\IEEEtitleabstractindextext{%
		\begin{abstract}
			Wearable sensor-based human activity recognition (HAR) has been a research focus in the field of ubiquitous and mobile computing for years. In recent years, many deep models have been applied to HAR problems. However, deep learning methods typically require a large amount of data for models to generalize well. Significant variances caused by different participants or diverse sensor devices limit the direct application of a pre-trained model to a subject or device that has not been seen before. To address these problems, we present an invariant feature learning framework (IFLF) that extracts common information shared across subjects and devices. IFLF incorporates two learning paradigms: 1) meta-learning to capture robust features across seen domains and adapt to an unseen one with similarity-based data selection; 2) multi-task learning to deal with data shortage and enhance overall performance via knowledge sharing among different subjects. Experiments demonstrated that IFLF is effective in handling both subject and device diversion across popular open datasets and an in-house dataset. It outperforms a baseline model of up to $40\%$ in test accuracy.
		\end{abstract}

		}
	
	\maketitle

	\section{Introduction}
\begin{figure*}
		
		\begin{subfigure}[t]{0.5\textwidth}
			\centering
			\includegraphics[height=1.2in]{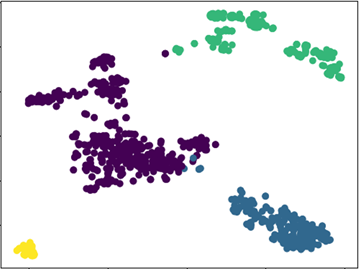}
			\includegraphics[height=1.2in]{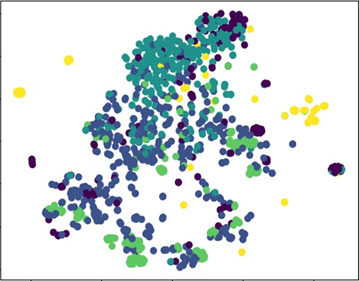}
			\caption{}
		\end{subfigure}%
		~ 
		\begin{subfigure}[t]{0.5\textwidth}
			\centering
			\includegraphics[height=1.2in]{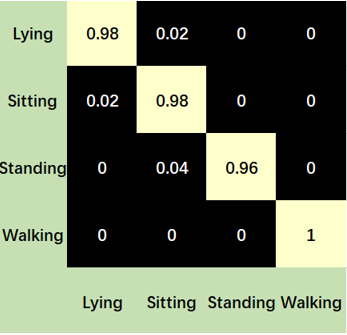}
			\includegraphics[height=1.2in]{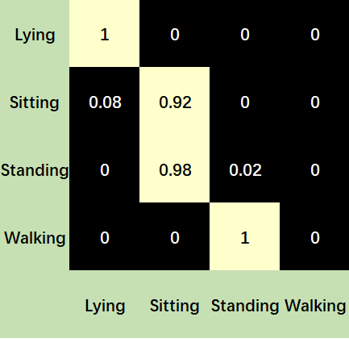}
			\caption{}
		\end{subfigure}
		\caption{Typical data variance problem caused by subject difference and device diversity. (a) depicts the impact of subject difference, where different colors in the t-SNE plots denote different activities. (b) shows the impact of device diversity. Left confusion matrix shows predictions on device seen to the model while right confusion matrix from a new unseen one. The prediction on unseen device's data is totally confused except for the `lying' activity.}
		\label{domain shifts}
	\end{figure*}

	Human activity recognition (HAR) is the foundation to realize remote health services and in-home mobility monitoring. Although deep learning has seen many successes in this field, training deep models often requires a large amount of sensory data that is not always available\revtr{\cite{kwon2020imutube}}. \revtr{For research ethics compliance, it often takes months to design study protocols, recruit volunteers and collect customized sensory datasets. At the same time, p}ublic inertial measurement unit (IMU) sensor datasets on HAR are typically collected by different groups of researchers following different experiment protocols, making them difficult to be used by others. The significant variability among human subjects and device types in data collection limits the direct reuse of data as well. Deep learning methods have poor generalization ability when testing data (target domain) differs from training data (source domain) due to device and subject heterogeneities (generally known as the {\it domain shift} problem). Fig. \ref{domain shifts} shows the effects of cross-domain data variability. Fig. 1(a) visualizes features from subjects that are seen to the deep model for HAR (left) and as held-out data to the model (right), respectively. The features are well clustered when subjects are seen to the model and are inseparable for the unseen subject even though she performs the same group of activities wearing the same sensor at the same location. Fig. 1(b) demonstrates the effect of device diversity. The data is collected when a person performs several activities with devices A and B attached to the same on-body locations. A deep learning model is trained with device A's data. We find that despite its high inference accuracy on the testing data from the same device, the accuracy drops drastically on device B's data. 
	
	In addressing the aforementioned domain shift problems, a pooling task model (PTM) that mixes data from different domains (e.g., subjects, devices)  will have low discriminative power as it ignores the dissimilarity among the domains. On the other hand, a model trained solely on data from a specific domain requires a lot of training data as it fails to take the advantage of the similarity among different sources. Since collecting and labeling sensory data with sufficient diversity is time-consuming, it is impractical to train a task-specific model for each new subject or device encountered from scratch. A few previous works have investigated domain shifts caused by device and subject diversity in HAR. In \cite{soleimani2019cross} and \cite{chen2019cross}, the problem is formulated as domain adaptation between a pair of participants or devices. However, in practice, HAR is rarely limited to transferring knowledge between a pair of domains, rather from a group of source domains (e.g., subjects, placements, or devices) to a target domain. Furthermore, unsupervised domain adaptation approaches trade-off their performance with data labeling efforts. For example, in \cite{soleimani2019cross}, the authors report an F1-score $\leq0.8$ in testing compared to 0.92 from \cite{ordonez2016deep} with supervised learning on the Opportunity dataset\cite{roggen2010collecting}.
	
	It is expected that despite their differences, sensory data of the same activity from different subjects or devices intrinsically share common characteristics. \revtr{A common assumption is the existence of shared representations among source and target domains. When tasks sampled from source domains can `cover' the representation space, linear predictors built upon feature extractor for each task is sufficient for good prediction\cite{kouw2019review,du2020few}. } Based on this assumption, we formulate the domain shift problem as a meta-learning problem with the aim to learn invariant features. By extracting features shared across domains and build task-specific layers for different source domains, the trained meta-model has better generalizability and can be adapted to a new target domain with few labeled data (a.k.a \textit{fast adaptation}). To further reduce the amount of labeled data, we devise a metric to qualify the similarity of activities from different domains. Such a metric allows us to selectively collect new labeled data for activities exhibit high domain shifts. We have evaluated IFLF using multiple public HAR datasets and one in-house datasets collected from older adults. Extensive experiments demonstrate that a test accuracy $\geq90\%$ can be achieved when only 10 seconds of sensory data per activity class is available from a target domain.
	
	The main contributions of this work are:
	\begin{itemize}
		\item We present an invariant feature learning framework that can handle various sources of domain shifts and demonstrate its superior performance with extensive experiments over multiple datasets. 
		\item IFLF alleviates data shortage through feature sharing from multiple source domains.
		\item A similarity metric is proposed to further reduce the amount of labeled data needed from a target domain for model adaptation.
	\end{itemize}
	
	The rest of paper is organized as follows. In Section 2 we discuss related work. Section 3 introduces the proposed invariant feature learning framework for HAR. We present the evaluation results on publicly available datasets and our own dataset in Section 4. Finally, Section 5 concludes the paper and lists future directions of study.
	
	\section{Related work}
	In this section, we first give an overview of HAR models. Next, we discuss two categories of approaches that address variations among different domains, namely, 1) domain adaptation and 2) domain-invariant feature learning.
	\subsection{Human Activity Recognition Models}
	Before the tide of deep learning, one popular approach to solve HAR problems is extracting a set of handcrafted features based on domain knowledge and training a shallow machine learning model\cite{yin2008sensor,anguita2012human,khan2014activity}. In \cite{khan2014activity}, two types of features are utilized, namely, time domain features (Mean, variance or standard deviation, energy, entropy, correlation between axes, signal magnitude area, tilt angle, and autoregressive coefficients) and frequency domain features (fast Fourier transform and discrete cosine transform coefficients). Accuracy of 99\% and 92\% is reported with a support vector machine model and a one-layer neural network model, respectively, built upon the features when classifying 16 activities. However, the effectiveness of handcrafted features can be highly activity specific.
	
	With deep learning, features can be learned from data automatically. A convolutional neural network is usually incorporated as part of the feature extractor. Deep models are reported to achieve state-of-the-art results on many popular open datasets \cite{ordonez2016deep,yang2015deep,peng2018aroma,yao2017deepsense}.  However, deep learning has its own limitations. It requires a large amount of data to train and is sensitive to domain shifts. \revtr{For example, the t-distributed stochastic neighbor embedding (t-SNE) visualization \cite{maaten2008visualizing} of 2D features, a form of non-linear embedding for high-dimension data, in Fig.1(a) was originally 128 dimensions extracted by DeepConvLSTM \cite{ordonez2016deep}. Its predictive accuracy drops dramatically when applied to unseen subjects and devices. DeepSense\cite{yao2017deepsense} is a neural network architecture that is robust to domain shifts by merging local interactions of different sensory modalities into global interactions. However, it requires multi-sensor modalities, long data windows to achieve good performance, and is sensitive to class imbalance, making it unsuitable for transient or highly dynamic activities.}
	
	\revtr{It should be noted that the proposed framework is model agnostic. In other words, we can incorporate any state-of-the-art deep learning architecture for HAR including DeepSense as the invariant feature extractor.}
	
	\subsection{Domain Adaptation}
	As a sub-category of transfer learning approaches, domain adaptation mitigates the problem when the training data used to learn a model has a different distribution from the data on which the model is applied\cite{patel2015visual}. Differed by the information available for the target task, domain adaptation approaches can be further divided into supervised\cite{motiian2017unified,tas2018cnn}, semi-supervised\cite{saito2019semi} and unsupervised domain adaptation\cite{soleimani2019cross,mathur2019mic2mic,akbari2019transferring,wang2018stratified}. 
	
	Previous work on sensor-based HAR mostly falls in the category of unsupervised domain adaptation. Three types of domain shifts have been considered, namely, subject difference, device diversity and sensor location divergence. In \cite{soleimani2019cross}, Soleimani and Nazerfard focus on mitigating subject differences when abundant unlabeled data is available in the target domain. A generative adversarial neural network (GAN) based solution is proposed to generate shared feature representation across a pair of source and target domains. Though not targeting HAR, the work  \cite{mathur2019mic2mic} is relevant to mitigate domain shifts due to device diversity. It utilizes a cycle-consistent generative adversarial network (CycleGAN) to transform target domain data to a source domain, and then apply a classifier trained on the source domain. In \cite{akbari2019transferring}, Akbari and Jafari propose a deep generative model to transfer knowledge between a labeled source sensor and an unlabeled target.  A mechanism to annotate pseudo labels for a target sensor was proposed by majority voting and intra-class correlation in \cite{wang2018stratified}. It is based on handcrafted features and a SVM model. Despite the popularity of unsupervised approaches, they trade-off the ability of learning invariant features that are robust to new domain with data labeling efforts. Thus, the performance tends to be noticeably inferior to supervised approaches. Also, the problem setup is limited to transfer knowledge between a pair of source and target domains which is limiting in practice. 
	
	\subsection{Domain-Invariant Feature Learning}
	Learning invariant features across different domains can facilitate a better generalization of a deep learning model. Specifically, meta-learning (a.k.a learning to learn \cite{schmidhuber1987evolutionary,bengio1990learning}) is one approach to achieve this goal. Model-agnostic meta-learning (MAML) \cite{finn2017model} introduces an episodic training paradigm with gradient-based parameter updating. It inspired meta-learning based HAR approaches in \cite{gui2018few,yu2018one,gong2019metasense}. The first two target computer vision tasks, \revtr{while MetaSense\cite{gong2019metasense} is designed for sensor-based HAR and thus is the most relevant to our work. In MetaSense, Gong et al. proposed to sample tasks both randomly within each source domain and across source domains. It achieves good performance with few-shot learning tests, but one limitation is the task sampling method requires each source domain to have the same number of classes. This assumption does not always hold especially when the activity set involves difficult or intensive motions. Both IFLF and MetaSense are meta-learning approaches but operate under distinctive assumptions. IFLF assumes that the source domains and the target domain share invariant features. In contrast, MAML and its variants such as MetaSense assume the existence of weights that are only a few gradient steps away from the optimal ones in every domain. These assumptions lead to different ways of updating models with data from the target domain: MAML and its variants update parameters of the whole model while IFLF only updates the task-specific layers. }
	
	Multi-task learning also helps to extract invariant features by learning the knowledge shared by different tasks (or domains). In \cite{sun2011new}, the authors propose a personalized shallow model for HAR, with a test accuracy between 63.9\% and 72.8\% on different experiment settings. It also considers a subject-level similarity as transfer factor that controls model parameter update in a gradient step. The work in  \cite{peng2018aroma,saeed2019multi} adopt deep learning methods. In \cite{peng2018aroma}, Peng et al. handle simple and complex activities with different task-specific layers on top of invariant features across them; whereas self-supervised learning is utilized through training an invariant feature extractor that is capable of extracting features behind various signal distortions in \cite{saeed2019multi}. There are 8 types of manually added signal distortions (random noise, scaled, rotated, negated, horizontally flipped, permuted, time-warped, and channel-shuffled) involved, but limited by the types of predefined signal distortions, it reaches an overall accuracy between 75.55\% and 88.55\% on 6 open datasets.
	
	To the best of our knowledge, ours is the first work to comprehensively deal with the domain shifts arising from multiple sources and data shortage problem. It differs from previous meta-learning approaches in that instead of updating all parameters of a meta-model in a gradient step, it trains a model in an alternating optimization manner \cite{kumar2012learning} to separate the task-specific and domain-invariant knowledge. In IFLF, a small amount of labeled data is required in meta-test step. Comparing to unsupervised domain adaptation approaches, doing so results in better classification accuracy at low costs and the ability to handle missing classes in source or target domains.
	
	\section{Method}
	Let the input and label spaces be $\mathcal{X}$ and  $\mathcal{Y}$, respectively. The target domain and the set of source domains are $\mathcal{D}_{tgt}=\{(x_{n},y_{n})\}_{n=1}^{M}$ and $\mathcal{D}_{src}=\{D_{1},D_{2},...,D_{K}\}$, respectively. $\mathcal{D}_{tgt}$ and $\mathcal{D}_{src}$ follow different distributions on the joint space $\mathcal{X}\times\mathcal{Y}$. A \textit{domain} $D_{k}=\{(x_{n}^{(k)},y_{n}^{(k)})\}_{n=1}^{N_{k}}$ corresponds to a source of variation, e.g., a subject or a device, where $N_{k}$ is the number of labeled data samples. In HAR, each \textit{task} is a multi-class classification problem that predicts the activity being performed from data sampled from the respective domain. The problem of meta-learning aims to learn well-generalized features from multiple source domains, and adapts the trained model to the target domain with small amount of labeled data. Since we assume the existence of domain-invariant features across the source and target domains, only the domain specific layers of the model need to be updated when applying to the target domain. 
	
	In this section, we present the detail of the proposed method. First, we illustrate the overall idea in Section \ref{sec:3.1}. Second, the detail of invariant feature learning will be introduced in Section \ref{sec:3.2}. Third, we explain in Section \ref{sec:3.4} the similarity-based fast adaptation. 
	\subsection{Overview}
	\label{sec:3.1}	
	The intuition behind IFLF is to learn two types of knowledge from multiple source domains: the shared features that can boost the generalization of a machine learning model, and the task-specific knowledge that provides the discriminative power within a specific domain. This is intrinsically reasonable for HAR problems: the task variations caused by different subjects or devices can be captured by task-specific parameters of IFLF. On the other hand, the signals of the same activity also have commonality, which can be embedded in an invariant feature representation that is shared across tasks. More importantly, such invariant features can also be transferred to a new HAR task to build a reliable model with very few data. To model the domain invariant features and task-specific ones respectively, IFLF is built upon a multi-task learning strategy where a task is associated with one of the source domains. 
	
\begin{figure}[h!]
		\centering
		\includegraphics[width=\linewidth]{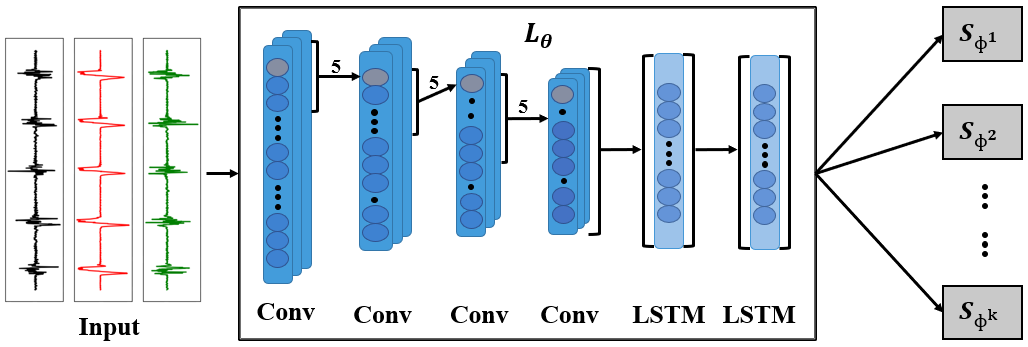}
		\caption{An example of IFLF model which employs 4 convolutional neural network (CNN) layers and 2 long short-term memory (LSTM) layers as $L_{\theta}$, and $K$ softmax layers as $S_{\phi^{k}}$ each corresponding to a domain in $\mathcal{D}_{src}$.}
		\label{fig:model}
	\end{figure}

	There are two key components in the proposed learning framework: 1) a feature extractor $L_{\theta}:\mathcal{X}\rightarrow\mathcal{Z}$, where $\mathcal{Z}$ is the feature space and $\theta$ denotes the parameters of $L$, and 2) a group of task-specific networks: $S_{\phi^{k}}: \mathcal{Z}\rightarrow \mathbb{R}^{C} $, where $k$ denotes the $k$-th task or domain, $\phi^{k}$ are the parameters of $k$-th task-specific layer $S^{k}$, and $C$ is the number of classes in $\mathcal{Y}$. Accordingly, the loss function is also composed of a feature extraction objective $\ell_{L}$ and a task-specific objective $\ell_{S^{k}}$, which will be detailed in Section \ref{sec:3.2}. The output of a task-specific network is given by $\hat{y}=softmax(S_{\phi^{k}}(L_{\theta}(x)))$, where $softmax(z)_{j}=e^{z_{j}}/\sum_{k=1}^{C}e^{z_{k}}$, for $j=1,...,C.$ IFLF is model-agnostic as $L_{\theta}$ and $S_{\phi^{k}}$ can be any reasonable neural network. An example neural network architecture of IFLF is shown in Fig.\ref{fig:model}. 
	
	In the training step, an IFLF model is meta-trained on $\mathcal{D}_{src}$ to obtain parameters $\theta$ and $\phi$. During testing in a target domain, the trained feature extractor network $L_{\theta}$ will be directly reused, while a new task-specific network need to be trained with task-specific data from $\mathcal{D}_{tgt}$. Algorithm 1 depicts the overall training process of IFLF, with learning rates hyperparameters $\alpha,\beta$. The algorithm optimizes $\theta$ and $\phi$ in an alternating way. 
	\begin{algorithm}[t]
		\caption{Invariant feature learning for domain adaptation}
		\begin{algorithmic}[1]
			\REQUIRE
			Source domains $\mathcal{D}_{src} = \{D_{k}\}_{k=1}^{K}$, hyperparameters $\alpha,\beta$
			\ENSURE
			IFLF model with parameter $\theta$ and $\phi$
			\State Random initialize $\theta,\phi=\{\phi^{1},\phi^{2},...,\phi^{K}\}$
			\REPEAT 
			\State Sample tasks $T=\{T_{1},T_{2},...,T_{K}\}$ over  $\mathcal{D}_{src}$;
			\State //Update $\phi^{k}$ with fixed $\theta$: 
			\FOR {$k$ is 1 to $K$}\\
			\hspace{\parindent} Freeze parameters of $\phi$ except $\phi^{k}$;\\
			\hspace{\parindent} $\phi^{k}\leftarrow\phi^{k}-\beta\nabla_{\phi^{k}}\ell_{S^{k}}(T_{k},\theta;\phi^{k})$;
			\ENDFOR 
			\State //Update $\theta$ with fixed $\phi$;
			\State  $\theta\leftarrow\theta-\alpha\nabla_{\theta}\ell_{L}(T,\phi;\theta)$;  
			
			\UNTIL convergence
		\end{algorithmic}
	\end{algorithm}
	
	\subsection{Invariant Feature Learning}
	\label{sec:3.2}
	To learn invariant features across source domains, one needs to consider three key factors: training strategy, feature extractor objective, and task-specific objective.
	\paragraph*{\textbf{Alternating Training}}If an IFLF model is trained by simply iterating among tasks sampled from $D_{1}$ to $D_{K}$, catastrophic forgetting may occur\cite{kirkpatrick2017overcoming}, namely, a model forgets previously learned tasks, and can only works properly on newly learned tasks. To avoid catastrophic forgetting, we adopt the alternating training strategy from \cite{kumar2012learning}, to update $L_{\theta}$ and $S_{\phi^{k}}$ separately. In each training epoch, we first freeze the parameters of the feature extractor layers, and update parameters of each task-specific layer with its respective data; then, we freeze parameters of the task-specific layers, and update the invariant feature extractor using all data from the previous step.
	\paragraph*{\textbf{Feature Extractor}}By the merit of multi-task learning, $L_{\theta}$ is designed to generalize well across domains through the sharing of  representations between related tasks \cite{ruder2017overview}. Despite its model-agnostic nature, we adopt a simple architecture described in \cite{ordonez2016deep} that is shown to be effective for HAR (See Fig. \ref{fig:model}). The network includes four convolutional neural network (CNN) layers and two long short-term memory (LSTM) layers. Because the application of convolution operator depends on the input dimension, we use a 1D kernel to convolve with one-dimensional temporal sequence (a.k.a the sensor signal) \cite{zeng2014convolutional}. 1D temporal CNNs are widely used in the area of sensor-based HAR (see \cite{wang2019deep} for a detailed survey), the combination of CNN and LSTM is beneficial for acquiring contextural knowledge and extracting meaningful features for time series data.
	
	The objective function $\ell_{L}$ works on multiple source domains to learn a domain invariant feature representation that clusters the features by their labels. It is defined as follows:
	
	\begin{equation}
		\label{eq:loss_l}
		\ell_{L} = \sum_{k=1}^{K}loss_{L}(T_{k},\phi;\theta),
	\end{equation} 
	where $loss_{L}$ is a loss function calculated on each $T_{k}$ with given $\theta$ and $\phi$. 
	Two types of loss functions are employed in this work. The first one is simply a categorical cross-entropy loss, defined as
	$loss_{L}=-\sum_{i=1}^{C}  y_{i}log(\hat{y}_{i})$
	on data from each task $k$. An IFLF model that uses cross-entropy in the loss term in equation (\ref{eq:loss_l}) is called a \textit{basic multi-task learning model} (BMTL). 
	
	In light of encouraging features to be locally clustered according to class regardless of the domain, we introduce the second type of loss function which utilizes triplet loss\cite{hoffer2015deep}. To calculate the triplet loss, one needs to sample $m$ triplets from raw data in $T_{k}$, and a triplet is $t=(x_{a},x_{p},x_{n})$. The corresponding output of a triplet in the feature space $\mathcal{Z}$ is $
	L_{\theta}(t)=(z_{a},z_{p},z_{n})$, where $x_{a}$ denotes the anchor sample, $x_{p}$ is the positive sample from the same class as $x_{a}$, and $x_{n}$ is a negative sample from class other than $x_{a}$. Here, the objective is to maximize the distance between $(z_{a},z_{n})$ and minimize the distance between $(z_{a},z_{p})$. Since it is difficult to determine a fixed threshold in a high dimensional space that separate data points into groups that are sufficiently close (and thus belong to the same class), a triplet loss is suitable for learning features that maximizes inter-class distances while minimizing intra-class distances. We compute the triplet loss as:
	\begin{equation}
		\label{eq:trip}
		loss_{L}(T_{k},\phi;\theta) =\sum_{i=1}^{m} \max\{0,{\left\|z_{a}^{i}-z_{p}^{i}\right\|\\}^{2}-{\left\|z_{a}^{i}-z_{n}^{i}\right\|\\}^{2}+\epsilon\},
	\end{equation}
	where $\epsilon$ is a margin enforced between positive and negative pairs\cite{schroff2015facenet}. An IFLF model with a triplet loss is called \textit{triplet multi-task learning model} (TMTL). Similar to BMTL, the loss function of TMTL is also calculated on each individual task. We then take the summation of losses over all source domains as the final objective function (\ref{eq:loss_l}).
	\paragraph*{\textbf{Task-specific Networks}} Under the assumption that if the shared feature generalizes well across all source domains, it will work on the target domain as well,  $L_{\theta}$ should be capable of exploring the entire latent space $\mathcal{Z}$ and extracting domain invariant feature. At the same time, a task-specific network $S_{\phi}^{k}$ should be simple to save the labor for fast adaptation, and be sparse to take only a subset (selected feature columns) from $\mathcal{Z}$ as its inputs. A lightweight architecture of a task-specific layer $S_{\phi^{k}}$ includes a fully connected layer with a softmax activation function. The task-specific objective function is defined as the sum of a categorical cross-entropy loss and an $\ell_1$-norm regularization term as follows,
	\begin{equation}
		\label{eq:loss_s}
		\ell_{S^{k}} = -\sum_{i=1}^{C}  y_{i}^{(k)}log(\hat{y}_{i}^{(k)}) + \mu |\phi^{k}|_{1},  
	\end{equation}
	where $\mu$ is a hyperparameter to control the sparsity. The regularization term imposes sparsity on the task-specific layers and helps mitigate overfitting. During meta-test, we can adapt the trained model to $\mathcal{D}_{tgt}$ by either initiating a new task-specific layer from scratch or updating the parameters of a selected $S_{\phi^{k}}$. An observation is that when features extracted by $L_{\theta}$ are well-clustered, we can randomly select one $\phi^{k}$ to conduct fast adaptation without much variance on the performance.
	
	\subsection{Similarity-based Fast Adaptation}
	\label{sec:3.4}
	Aiming at further reducing the amount of labeled data required from the target domain for fast adaptation, a metric helps to identify the similarity or dissimilarity of motion patterns is required. \revtr{We assume that if  similar patterns are observed on an activity among all source domains, it is highly likely that the same activity in the target domain follows the same pattern as well.} To quantify the similarity of two sensor signals, we propose a similarity metric in equation (\ref{eq:similar}), which is calculated between data from the same activity across all source domains.
	\begin{equation}
		\label{eq:similar}
		similarity_{i,j}^{c}=\sum_{i=1}^{K}\sum_{j=1}^{K}\dfrac{cov(p^{i},p^{j})}{\sigma_{p^{i}}\sigma_{p^{j}}},
	\end{equation}
	where $\sigma$ is standard deviation,  $(p^{i},p^{j})=$DTW$(x_{i},x_{j})$ is the pair of warped signals from sensor readings $x_{i}$ and $x_{j}$ from subject $i$ and $j$, $c$ is the activity class and $cov(\cdot,\cdot)$ is the covariance. Dynamic time warping (DTW)\cite{sakoe1978dynamic} calculates the best match between two temporal sequences, which may vary in speed. Here we use it to align raw sensory readings to mitigate time shifts and speed divergence. The Pearson correlation coefficient calculated on a pair of warped signals in equation (\ref{eq:similar}) provides a normalized similarity score that measures if activity $c$ is performed similarly between a pair of participants. Data needs to be pre-processed (e.g., interpolated, noise filtered and normalized) before feeding to DTW. To eliminate the impact of misaligned sensor axis, we use the magnitude per sensor (e.g., an accelerometer or a gyroscope) as input to the similarity calculation.
	
	\revtr{Consider measurements from two sensors attached to two subjects (Subject 1 and Subject 2) performing the same activity. If the warped distance of the resulting measurements is small, this implies that the movement patterns are similar between the two subjects for the activity (despite possible differences in pace). Therefore, we can safely substitute Subject 1's data with that of Subject 2 and vice versa. After an IFLF model is trained, we no longer need to obtain labeled data from every class in $\mathcal{D}_{tgt}$. For activities that are considered similar across all source domains, we simply sample from the corresponding activity data in the source domains. These samples together with labeled data for remaining classes from the target domain are then used in updating the parameters in the task-specific layers while keeping the feature extraction layers unchanged during gradient descent.} In the experiments, we find that a threshold $\ge0.8$ is suitable for distinguishing whether an activity is performed similarly among different subjects using the similarity measure defined in equation (\ref{eq:similar}).
	
	\section{Performance Evaluation}
	As our research mainly focuses on assessing the mobility status of older adults with IMU sensory data, we choose to conduct the experiments on open datasets and our own dataset on locomotion or lower limb exercises. During data collection, sensors are mainly attached to the trunk or lower limbs of participants. Nevertheless, the method proposed in this work is generic and can be applied to other types of activities and sensor placements. 
	
	\subsection{Datasets}
	\label{chapter4.1}
	We consider three publicly available datasets to cover a wide variety of device types, data collection protocols, and activity classes to be recognized, \revtr{and one in-house dataset that contains measurement data from multiple IMU sensors of different vendors on older patients.} Important aspects of the datasets are summarized in Table \ref{tab:datasets} with brief descriptions listed below. 
	
\begin{figure}[h!]
		\centering
		\includegraphics[width=0.8\linewidth]{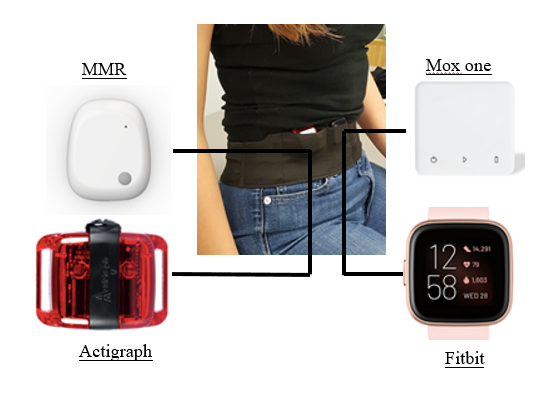}
		\caption{The sensors and their placement when collecting data for the MobilityAI dataset.}
		\label{fig:mobilityAI}
	\end{figure}

	\begin{enumerate}[label=(\roman*)]
		\begin{table*}[t]
			\caption{\label{tab:datasets} Summary of datasets used in our experiment. These datasets are selected based on the diversity of participants and popularity in the HAR research field. The number of activities listed here are locomotion only, and missing classes denotes the number of activity classes per subject may vary. Length of trials is reported as an average on subject. Further details on data pre-processing and information of each dataset is discussed in Section \ref{chapter4.1}.}
			\resizebox{\linewidth}{!}{%
				\begin{tabular}{lllllllll}
					\hline
					\multirow{ 2}{*}{} Dataset     & Sampling rate  & \#Sensors & Placement             & \#Activities & \#Subjects & Missing classes & Balanced &Length (in mins)\\
					
					\hline
					PAMAP2      & 100 Hz            & 1             & dominant side's ankle  & 8            & 8          & Yes   & No       & 59.67                                      \\
					USCHAD      & 100 Hz           & 1             & right hip     &10           & 14         & No      & Yes    & 29.54                                      \\
					WISDM       & 20 Hz             & 1             & pant pocket           & 7            & 51         & Yes    &Yes  &53.59                                          \\
					MobilityAI  & 50 Hz             & 4             & waist                 & 4            &25         &Yes  &No &15.29\\ 
					\hline
			\end{tabular}}
		\end{table*}
		\item\textbf{PAMAP2.} The Physical Activity Monitoring version 2 (PAMAP2) \cite{reiss2012introducing} is a dataset collected from one dominant ankle sensor (accelerometer and gyroscope) for 8 different activities, i.e., lying, sitting, walking, running, cycling, nordic walking, ascending stairs and descending stairs. Eight participants performed these activities freely without time constraints and have the option to skip some activities. Thus, there exist missing classes in some participants' data as well as unbalanced data samples across the classes. During data collection, sensors of the same model are instrumented on different subjects running at a sampling rate of 100Hz.
		\item\textbf{USCHAD.} This dataset \cite{zhang2012usc} is collected from 14 participants performing 10 types of locomotions (i.e., walking forward, walking left, walking right, walking upstairs, walking downstair, running forward, jumping up, sitting, standing and sleeping). All activities are performed by each subject in a controlled environment. A sensor (acclerometer and gyroscope) with a sampling rate of 100Hz is attached to the right hip of each participant. 5 data trials per activity were collected per participant, and the duration of each data trial varies. 
		\item\textbf{WISDM.} This dataset \cite{weiss2019smartphone} contains a large number of subjects. Raw accelerometer and gyroscope data have been collected from a smartphone in each participant's pant pocket at a rate of 20Hz. There are a total of 51 test subjects performing 7 locomotion activities (i.e., walking, jogging, stairs, sitting, standing, kicking soccer ball, playing tennis) for 3 minutes apiece to get equal class distribution. 
		
		\item\textbf{MobilityAI.} The mobility analysis by artificial intelligence (MobilityAI) dataset is collected from 25 in-hospital patients whose ages are $\geq65$. \revtr{The objective of collecting such a dataset is to monitor patients' mobility status during their hospital stay, and to quantify the effectiveness of an early mobilization protocol. To identify the most suitable device,} four IMU sensors from different vendors are attached to patients' waists using elastic bands as shown in Fig. \ref{fig:mobilityAI}. Each subject performs four activities (lying for 5 minutes, sitting for 5 minutes, standing for 5 minutes and 20 meters walking) for mobility status assessment. The four devices utilized are MetaMotionR  \cite{MMR}, Fitbit Versa \cite{FB}, Mox One \cite{Mox} and Actigraph \cite{Acti}. All sensors are set to have a 50Hz sampling rate, and only accelerometer readings are captured. Due to data outage and the limited functional mobility of some participants, there exist missing classes in the dataset.
	\end{enumerate}
	
	\subsection{Implementation and Evaluation Process}
	In addition to BMTL and TMTL discussed in the previous section, we have also implemented four baseline models: a single-task learning model (STL), a pooling task model (PTM) \revtr{and  MetaSense\cite{gong2019metasense}}. STL is trained solely on the target domain for comparison with domain adaptation, whereas PTM is trained with mixed training data from all source domains, to highlight the domain shift problem.
	
	\paragraph*{\textbf{Data Preparation}} Although deep neural networks can directly learn useful features from raw data \cite{yang2015deep}, data preprocessing such as interpolation, noise filtering, normalization, and the division of sliding windows are still needed. A Butterworth low-pass filter \cite{butterworth1930theory} with a cut-off frequency of 10Hz is employed to remove high frequency noise from interpolated data. After low-pass filtering, we normalize the data, \revtr{calculate similarity metrics,} and then segment it into sliding windows with a fixed length of 2 seconds with 80\% overlapping \revtr{for all datasets}. To eliminate the impact of different orientations of sensors in MobilityAI, we rotate the orientations of Actigraph, Mox one and Fitbit 3-axis accelerometers to be aligned with that of MetaMotionR. 
	
	\paragraph*{\textbf{Implementation}}
	\revtr{The implementation of feature extractor follows DeepConvLSTM \cite{ordonez2016deep} for IFLF, STL and PTM models. It includes four layers of 1D CNN and} two LSTM layers with 128 hidden units and a drop-out rate of 0.25 to prevent overfitting\cite{srivastava2014dropout}. \revtr{The CNN layers have 64 channels with kernel size 5 and stride 1. For a fair comparison with MetaSense, we also implement TMTL use the same network architecture as in \cite{gong2019metasense} based on an open  source implementation of MAML \cite{MAML_Pytorch}. The feature extractor network has five CNN layers and two fully-connected layers, including 128 and 64 hidden units respectively. The reason for not including LSTM layers to MetaSense is two-fold: 1) MAML based approaches require a 2-steps update on layer parameters, keeping intermediate variable for calculated gradients without actual updating. But existing deep learning libraries combines gradient calculation with backpropogation for recurrent neural network parameters, leaving no API for gradient calculation only, and 2) as a model agnostic approach, it is interesting to investigate the performance of IFLF without LSTM as well. Tasks in MetaSense are sampled both within and cross different source domains, keeping activity labels consistent across all tasks. }
	
	For STL and PTM, the output layer corresponds to a fully-connected layer with a softmax activation function. Both models are trained with a RMSProp optimizor\cite{bengio2015rmsprop} at a learning rate of $10^{-3}$ and a decay factor of $p=0.9$. The maximum iteration number is set to be 100. IFLF models utilize the aforementioned network structure as $L_{\theta}$, the number of $S_{\phi^{k}}$ branches is determined by the number of source domains, and \revtr{each $S_{\phi^{k}}$ may have a different output shape depending on the number of classes.} IFLF models are trained with an Adam\cite{kingma2014adam} optimizer at a learning rate of $10^{-4},\beta_{1}=0.9,\beta_{2}=0.999$, and hyper-parameters $\mu=0.8$. The batch size is set to 100 and the maximum number of training epochs is 30 with early-stopping. In each epoch, TMTL samples $m$ pairs of ($x_{a},x_{p}$) and $m$ pairs of ($x_{a},x_{n}$) to form $m^{2}$ ($x_{a},x_{p},x_{n}$) triplets from each source domain as task $T_{k}$. Similarly, $n$ pairs of ($x_{a},x_{p}$) and $n$ pairs of ($x_{a},x_{n}$) are sampled to form $n^{2}$ ($x_{a},x_{p},x_{n}$) triplets as validation set($m>n$). In most experiments, we set $m=100$, $n=10$ and hyperparameter $\epsilon=0.4$. The hyper-parameters and optimizer of each model are the same across all datasets.

	\paragraph*{\textbf{Evaluation Process}}Leave-one-domain-out evaluations are conducted on all datasets. \revtr{Under different problem settings, a domain can be a subject, a sensor device or a combination of subject and device. In each experiment, a target domain was randomly selected. Similarity calculation and model meta-training utilize data from source domains only.} 
	
	After the IFLF model is trained, we randomly sample a fixed test set from $\mathcal{D}_{tgt}$, and randomly select $i$ data windows (of 2s length) per class from the remaining data as the training set \revtr{to update a trained MetaSense model and an arbitrary $S_{\phi^{k}}$ layer of the IFLF model. This process is also called $i$-shot learning}. An STL model is also trained on this training set and test accuracy is recorded for each model by gradually increasing $i$ from 1 to 100. Each evaluation is repeated 5 times and the average is reported. 
	
	\subsection{Results}
	\label{sec:Results}
	\subsubsection{Evaluate with various domain shifts}
\begin{figure}[h!]
		\begin{minipage}{\linewidth}
			\centering
			\includegraphics[width=0.48\linewidth]{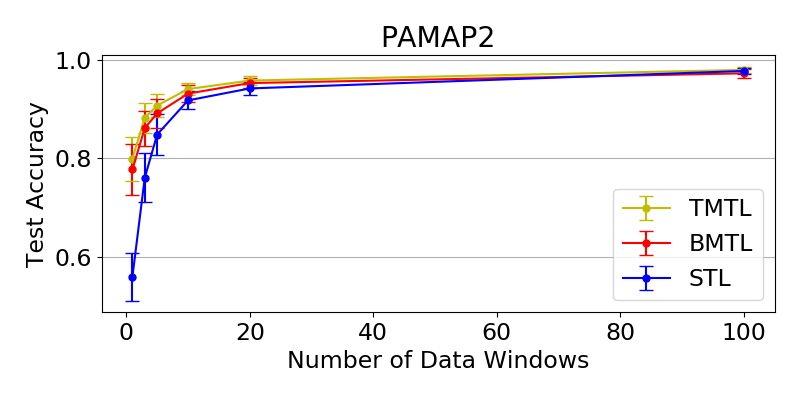}
			\includegraphics[width=0.48\linewidth]{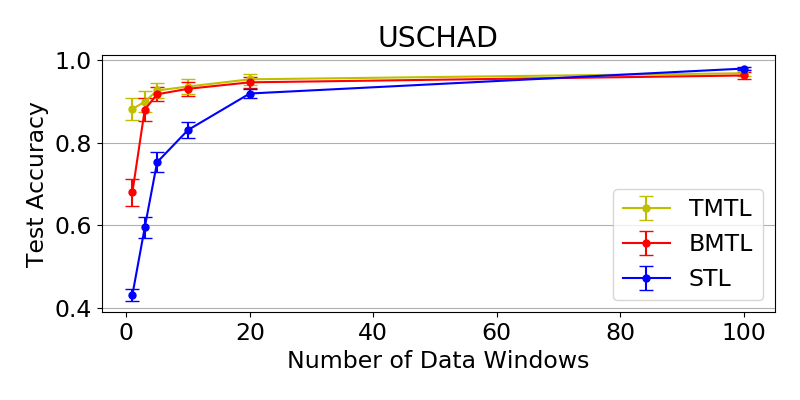}
			\includegraphics[width=0.48\linewidth]{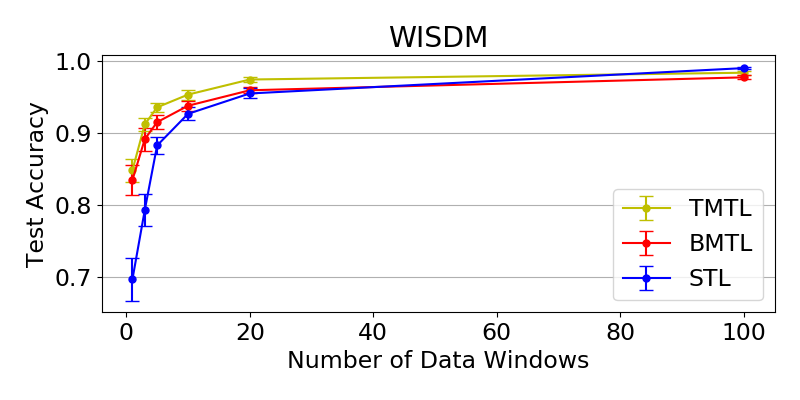}
			\includegraphics[width=0.48\linewidth]{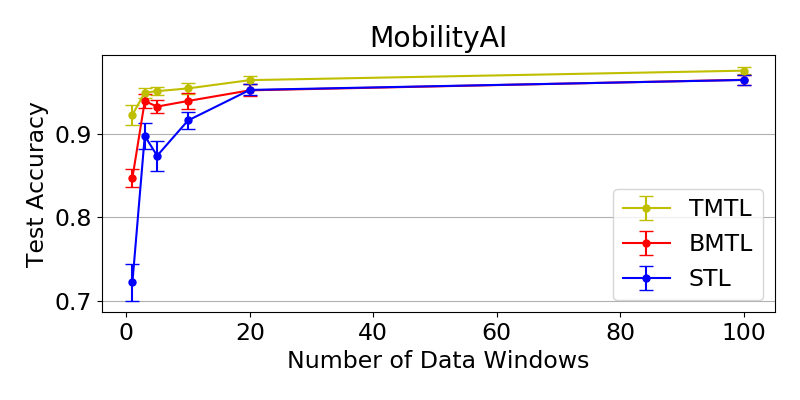}
		\end{minipage}
		
		\caption{Evaluation on subject difference across all datasets when gradually increase the number of data windows per activity class from $\mathcal{D}_{tgt}$. The test accuracy and standard error are averaged across different subjects in leave-one-out experiments.}
		\label{tab:subject difference}
	\end{figure}

	In order to evaluate IFLF's capability to handle domain shifts, three types of experiments are conducted on subject difference, device diversity and their combination (with both unseen participant and device).
	\paragraph*{\textbf{Subject Difference}}Fig. \ref{tab:subject difference} shows the averaged test accuracy with standard errors across all datasets. The average is computed over all subjects of each dataset.
	As demonstrated in Fig. \ref{tab:subject difference}, in terms of the overall test accuracy among the four models follows, BMTL is better than STL, and TMTL is better than BMTL when few data samples are available from $\mathcal{D}_{tgt}$. However, since STL is solely trained on the target domain, when $i$ is sufficiently large, its accuracy approaches 100\% and tends to be better than both IFLF models. We also observe that with different subjects as the target domain, the converging rate of STL is dramatically different, an indication of subject differences. 
	
\begin{figure}[h!]
		\begin{minipage}{\linewidth}
			\includegraphics[width=0.49\linewidth]{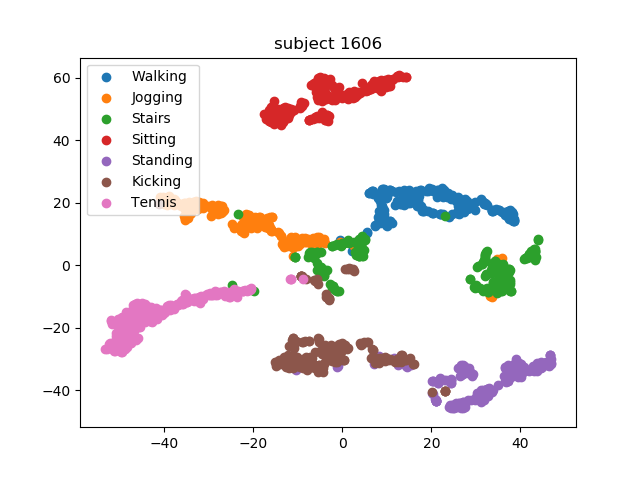}
			\includegraphics[width=0.49\linewidth]{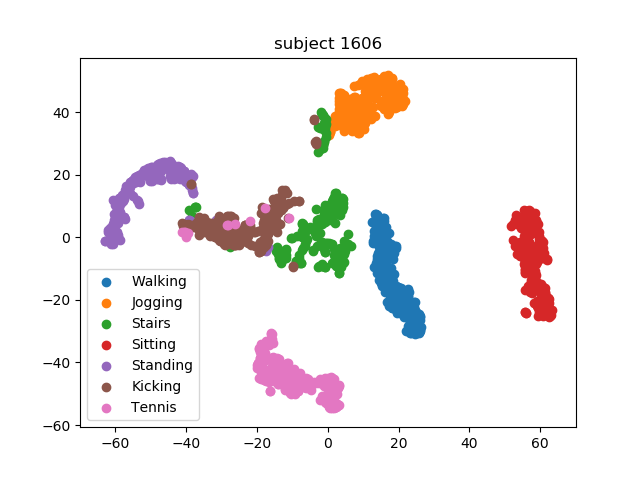}
			\subcaption{PTM models}
			\includegraphics[width=0.49\linewidth]{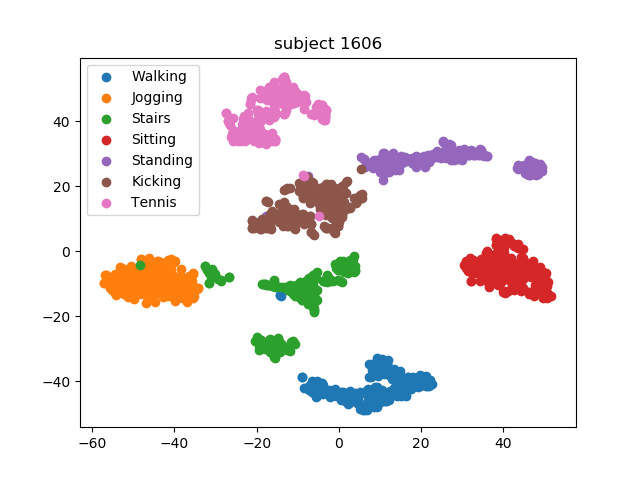}
			\includegraphics[width=0.49\linewidth]{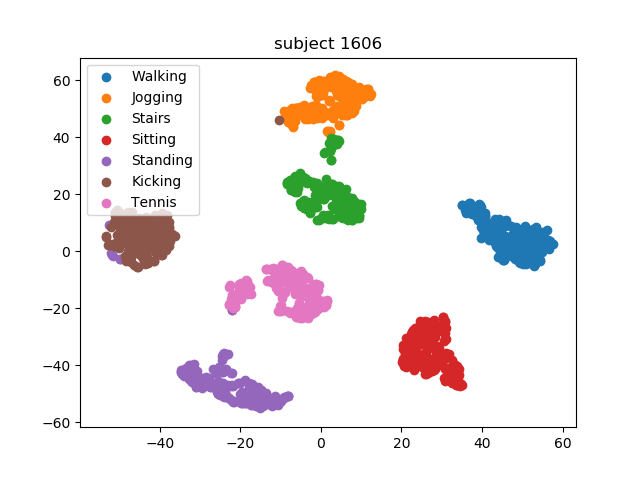}
			\subcaption{BMTL and TMTL model}
		\end{minipage}
		
		\caption{t-SNE visualization of the learned representations. We visualize the features from each model with the output of $L_{\theta}$ by projection them on 2D space. This example is generated by different models on WISDM dataset. (a) are PTM models; (b) left is BMTL and (b) right is TMTL.}
		\label{fig:tsne}
	\end{figure}

	To visualize the features produced by the methods, Fig. \ref{fig:tsne} shows the t-SNE of unseen subject's features produced by each model \cite{maaten2008visualizing}. When generating these plots, we randomly pick a subject as the target domain. The comparison is made among a PTM model, leaving this subject out in the BMTL model and TMTL model. To demonstrate the generality of the results, plots from two PTM models are presented with different random splits of the training set for each source domain. Although the figures are generated from the WISDM dataset, similar observations can be made for other datasets.
	
	From Fig.\ref{fig:tsne}, it is clear that the features for different activities generated by PTMs are entangled. BMTL improves the separation among activities to some extent, whereas TMTL generates a set of features with clear clustered structures and large margins. This also explained why the fast adaptation can be made on any trained $S_{\phi^{k}}$ layer of BMTL and TMTL. As the features are well separated, different choices of the task-specific layer for parameter update have little impact on the performance. However, as the amount of data is quite limited in the fast adaptation step, initializing $L_{\theta}$ randomly will impair the performance of IFLF. But even in this case, we find the IFLF models still work better than STL.
	
\begin{figure}[h!]
		\centering
		\begin{minipage}[b]{0.32\linewidth}
			\includegraphics[width=\linewidth]{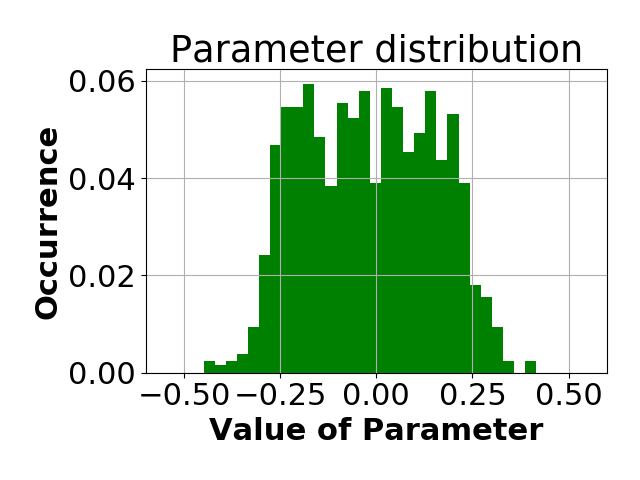}
			\subcaption{}
			
		\end{minipage}
		\begin{minipage}[b]{0.32\linewidth}
			\includegraphics[width=\linewidth]{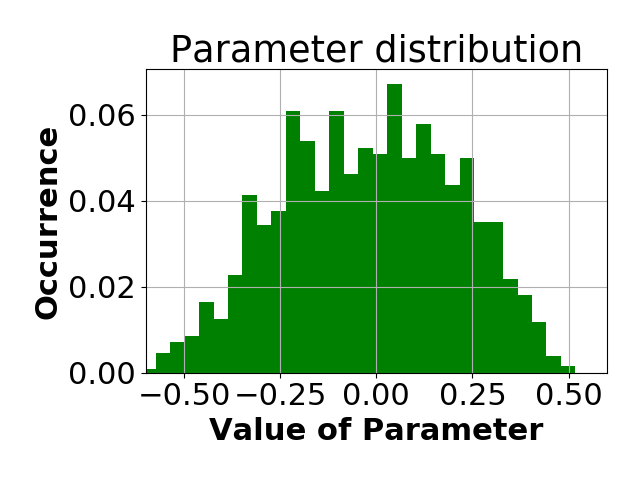}
			\subcaption{}
			
		\end{minipage}
		\begin{minipage}[b]{0.32\linewidth}
			\includegraphics[width=\linewidth]{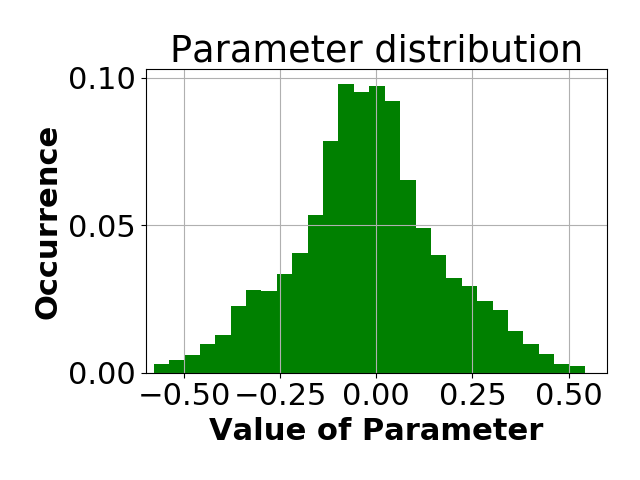}
			\subcaption{}
			
		\end{minipage}
		\caption{Comparison of the parameter $\phi$'s distribution for different models. From (a) to (c) are: STL, BMTL, TMTL. X-axis is the value of parameter, y-axis is the normalized occurrence.
		}
		\label{fig:classifier}
	\end{figure}

	To further illustrate the advantage of TMTL over the other two models, the sparsity of the parameters ($\phi$) of the learned task-specific layer is compared. In Fig. \ref{fig:classifier}, we plot the distribution of $\phi$ in the three models trained on WISDM dataset. 
	
	As shown in Fig. \ref{fig:classifier}, the parameters in STL are roughly a uniformly distributed between -0.4 and 0.4. In comparison, the parameters of BMTL follow a zero mean Gaussian distribution but with a large variance. Lastly, the majority of TMTL parameters are centered around 0 with a noticeably smaller variance (than that of BMTL). The sparsity of task-specific layer's parameters indicates the easiness of separating the generated feature representations. 
	\paragraph*{\textbf{Device Diversity}}Similar to subject difference experiment, IFLF can also tackle domain shift problems caused by device diversity. To understand the behavior of STL, BMTL, and TMTL to handle device diversity, we consider data from different sensor devices attached to the same subject at the same on-body position. Since only the MobilityAI dataset has such characteristics, it is utilized in the subsequent experiments. 
	
	In following experiments, data from Actigraph, Fitbit, Mox One constitute the source domains, while MMR device's data is selected as $D_{tgt}$. One data window per activity sampled from $D_{tgt}$ is utilized. To demonstrate the presence of domain shifts between different devices, besides the STL, BMTL and TMTL models, we further present the confusion matrix from PTM trained with data from Actigraph, Fitbit and Mox one but tested on MMR.
	\begin{table}[t]
		\centering
		\caption{\label{tab:confusion_matrix}The experiment on sensor diversity. Confusion matrices are generated with only 1 data window per activity involved.}
		\subcaption{The confusion matrix of the single task model trained on MetaMotionR sensor data.}
		\begin{tabular}{lllll}
			\hline
			& Lying & Sitting & Standing & Walking \\ \hline
			Lying    & 0.98  & 0.02    & 0        & 0       \\ \hline
			Sitting  & 0.02  & 0.98    & 0        & 0       \\ \hline
			Standing & 0     & 0.35    & 0.65     & 0       \\ \hline
			Walking  & 0     & 0       & 1        & 0       \\ \hline
			
		\end{tabular}\\
		\vspace{3mm}
		
		\subcaption{The confusion matrix of the BMTL model fast adapted with MetaMotionR data.}
		\begin{tabular}{lllll}
			\hline
			& Lying & Sitting & Standing & Walking \\ \hline
			Lying    & 0.98  & 0.02    & 0        & 0       \\ \hline
			Sitting  & 0.02  & 0.98    & 0        & 0       \\ \hline
			Standing & 0     & 0.42    & 0.58    & 0       \\ \hline
			Walking  & 0     & 0       & 0        & 1      \\ \hline
		\end{tabular}\\
		\vspace{3mm}
		
		\subcaption{The confusion matrix of the TMTL model fast adapted with MetaMotionR data.}
		\begin{tabular}{lllll}
			\hline
			& Lying & Sitting & Standing & Walking \\ \hline
			Lying    & 0.98  & 0.02    & 0        & 0       \\ \hline
			Sitting  & 0.08  & 0.90    & 0.02        & 0       \\ \hline
			Standing & 0     & 0.04    & 0.96    & 0       \\ \hline
			Walking  & 0     & 0       & 0        & 1      \\ \hline
		\end{tabular}\\
		\vspace{3mm}
		
		\subcaption{\label{2d}The confusion matrix of the PTM model trained without MetaMotionR and tested on MetaMotionR data directly.}
		\begin{tabular}{lllll}
			\hline
			& Lying & Sitting & Standing & Walking \\ \hline
			Lying    & 1  & 0.02    & 0        & 0       \\ \hline
			Sitting  & 0  & 0.99    & 0.01        & 0       \\ \hline
			Standing & 0.01     & 0.94    & 0.05   & 0       \\ \hline
			Walking  & 0     & 0.48       & 0.52        & 0      \\ \hline
		\end{tabular}
	\end{table}
	From the confusion matrices in Table \ref{tab:confusion_matrix}, it is clear that BMTL benefiting from invariant feature learning outperformed STL, while TMTL rarely misclassifies any activity in the dataset. Table II(d) shows that PTM is incapable of learning robust features that generalize well to the target domain.
	
\begin{figure}[h!]
		\begin{minipage}{\linewidth}
			\includegraphics[width=0.49\linewidth]{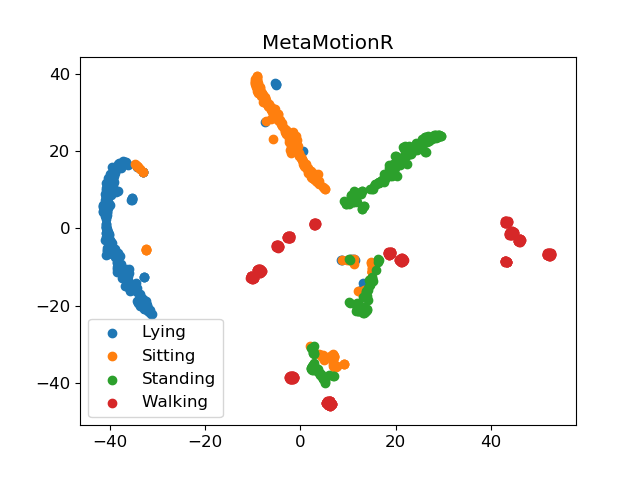}
			\includegraphics[width=0.49\linewidth]{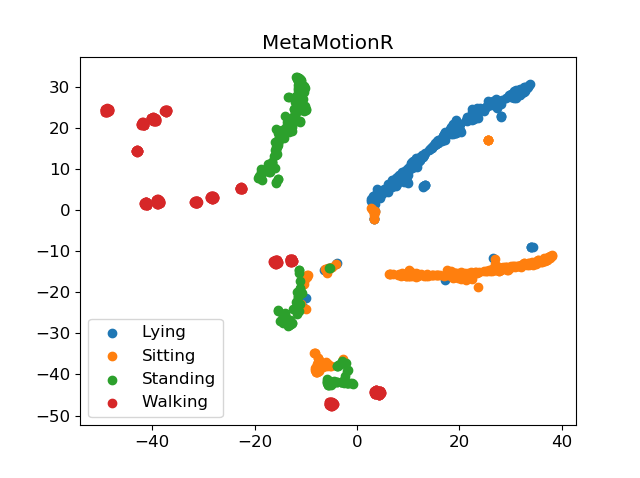}
			\subcaption{PTM models}
			\includegraphics[width=0.49\linewidth]{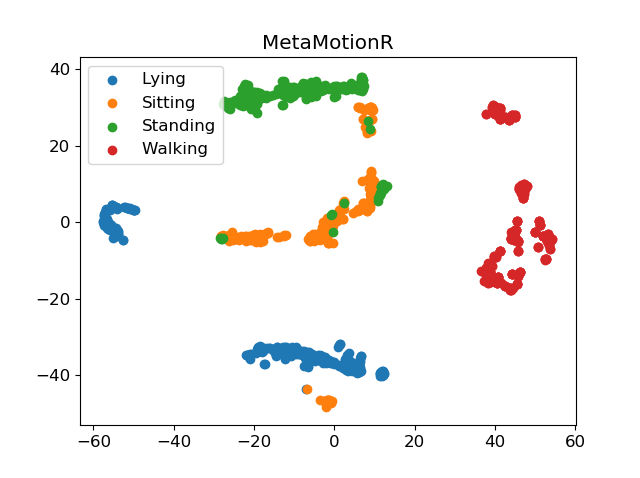}
			\includegraphics[width=0.49\linewidth]{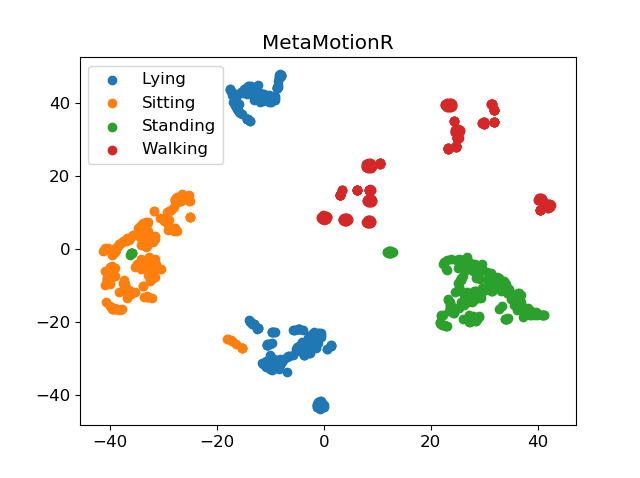}
			\subcaption{BMTL and TMTL model}
		\end{minipage}
		
		\caption{t-SNE visualization of the learned representations. This example is generated by different models on MobilityAI dataset. (a) are PTM models; (b) left is BMTL and (b) right is TMTL.}
		\label{fig:dv_tsne}
	\end{figure}

\begin{figure}[h!]
		\centering
		\begin{minipage}[b]{0.32\linewidth}
			\includegraphics[width=\linewidth]{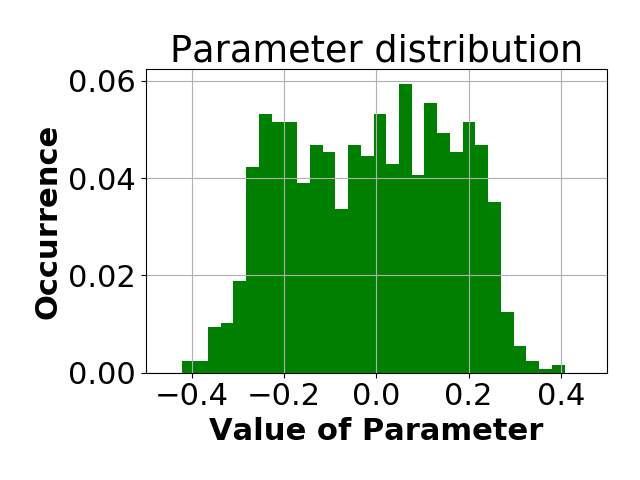}
			\subcaption{}
			
		\end{minipage}
		\begin{minipage}[b]{0.32\linewidth}
			\includegraphics[width=\linewidth]{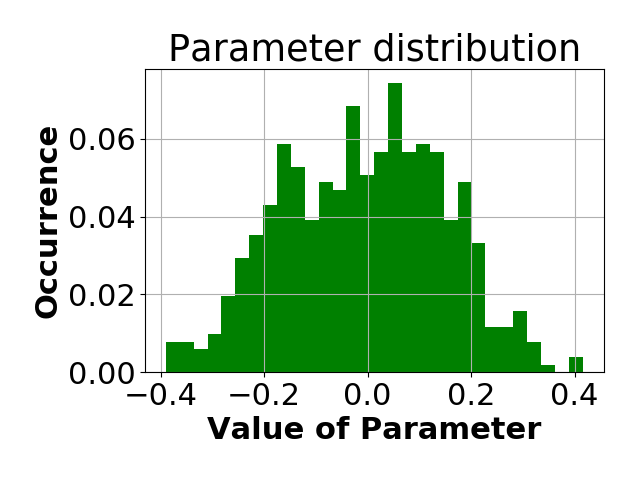}
			\subcaption{}
			
		\end{minipage}
		\begin{minipage}[b]{0.32\linewidth}
			\includegraphics[width=\linewidth]{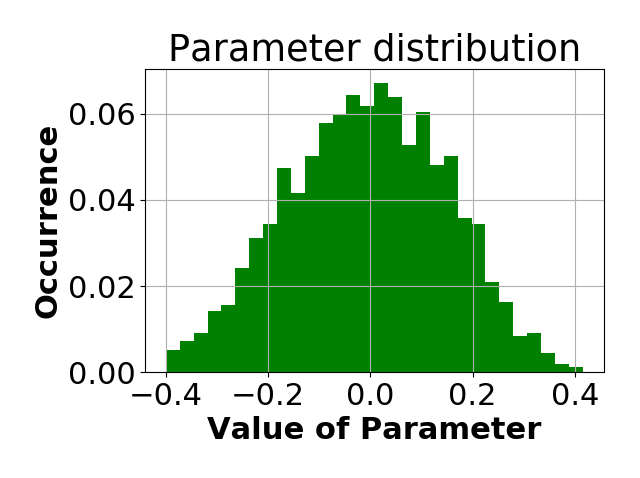}
			\subcaption{}
			
		\end{minipage}
		\caption{Comparison of the parameter $\phi$'s distribution for different models. From (a) to (c) are: STL, BMTL, TMTL.
		}
		\label{fig:dv_classifier}
	\end{figure}

	Fig. \ref{fig:dv_tsne} visualizes features extracted by PTM (the two plots in Fig. 7(a) are generated from different random splits of the training data), BMTL and TMTL models.
	Similar to the case of subject differences, we observe that the PTM models fail to extract separable features from MetaMotionR data while BMTL does a better job, but the resulting features are still not well-clustered. In contrast, TMTL gives rise to features with clear boundaries in the feature space and clustered structures. Fig. \ref{fig:dv_classifier} depicts the distribution of parameters $\phi$ in the task specific layers of each model. Similar to Fig. \ref{fig:classifier}, we observe that TMTL has the most sparsity, followed by BMTL, whereas STL leads to the least sparsity. The evaluation on device diversity further demonstrates the IFLF models' capability of learning invariant features across domains.
	
	\paragraph*{\textbf{Both subject and device are unseen}}Encouraged by the promising results on the subject difference and device diversity experiments, we further evaluate situations where both device and subject are unseen to the model. We randomly selected a subset from the MobilityAI dataset, which includes 8 subjects with waist attached sensors. As each subject has 4 sensor devices attached, the total combination of subject and sensor is 32. A pair of BMTL and TMTL models are trained on data from 5 participants each with 3 sensors attached, and evaluated on the 4th device data collected from participants other than the five in the training data. We conduct leave-one-out experiment on both subject and device. To keep brief, we only present results from one division of subjects as the other cases are quite similar. Specifically, the meta-training set includes data from Subject 1 to 5. The test data is from Subject 6, 7 and 8. 
	\begin{figure}[h!]
		\begin{minipage}[b]{\linewidth}
			\includegraphics[width=0.49\linewidth]{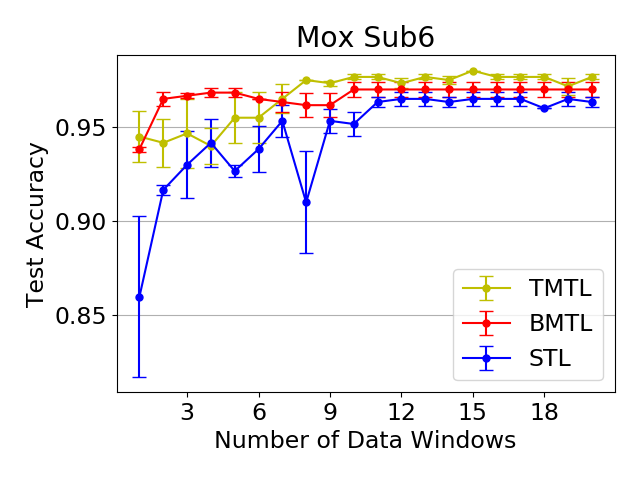}
			\includegraphics[width=0.49\linewidth]{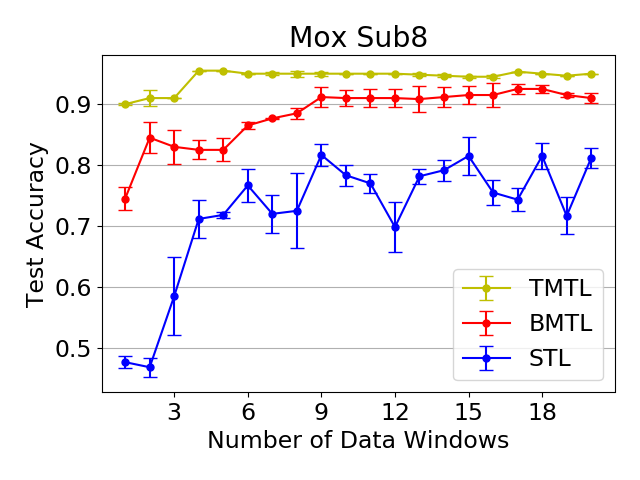}
			\includegraphics[width=0.49\linewidth]{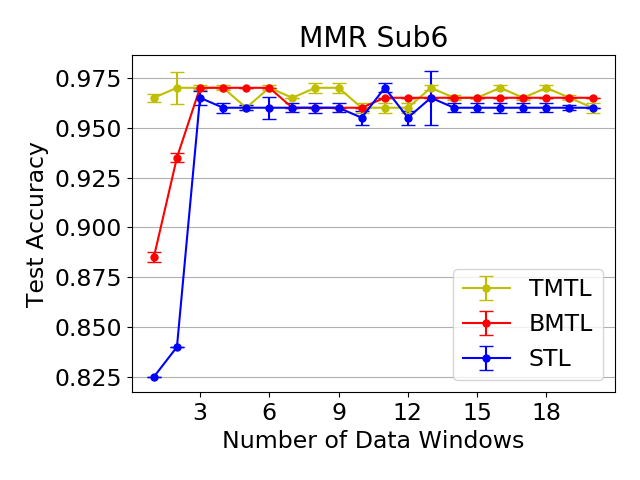}
			\includegraphics[width=0.49\linewidth]{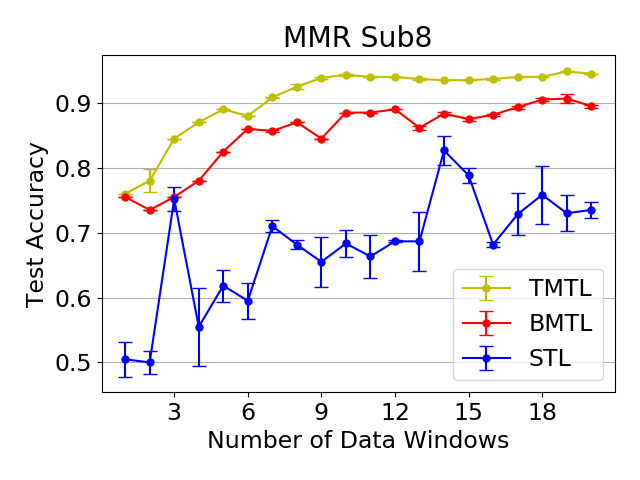}
		\end{minipage}
		\caption{Combinations of different sources of diversity (Both the test subject and the test device are not included in the training data for IFLF models). }
		\label{fig:combination}
	\end{figure}
	Fig. \ref{fig:combination} compares the performance when both the subject and device are unseen to IFLF models. In this case, both BMTL and TMTL have better performance than STL, especially when the amount of labeled data from the target domain is small. Compared to Fig. \ref{tab:subject difference}, we observe large gaps between TMTL and STL test accuracy. This can be attributed to the significant diversity among different devices. Furthermore, by comparing the results for Subject 6 and 8 for the same device (e.g., Mox One or MMR), we find that Subject 8 appears to have larger differences from subjects in the training set than Subject 6. Despite such differences, TMTL consistently outperforms BMTL and STL. Meanwhile, large standard errors are observed with STL as it is heavily dependent on the training set. 
	
	In summary, IFLF is a general method to capture invariant features, and works well regardless of the cause of domain shifts. 
	\subsubsection{Comparison with MetaSense}
	\revtr{To this end, we conclude TMTL outperforms STL, BMTL in handling domain shifts caused by subject and device diversity. Next, we present the comparison between TMTL without LSTM and the state-of-the-art meta-learning model for HAR, MetaSense. With PAMAP2, WISDM and USCHAD which only contain data from a single sensor, we compare the performance of these two models on mitigating domain shifts caused by subject differences. With MobilityAI dataset that include data from multiple subjects and devices, the performances of the two models when both subject and device are unseen are compared.}
	\begin{figure}[h!]
		\begin{minipage}[b]{\linewidth}
			\includegraphics[width=0.49\linewidth]{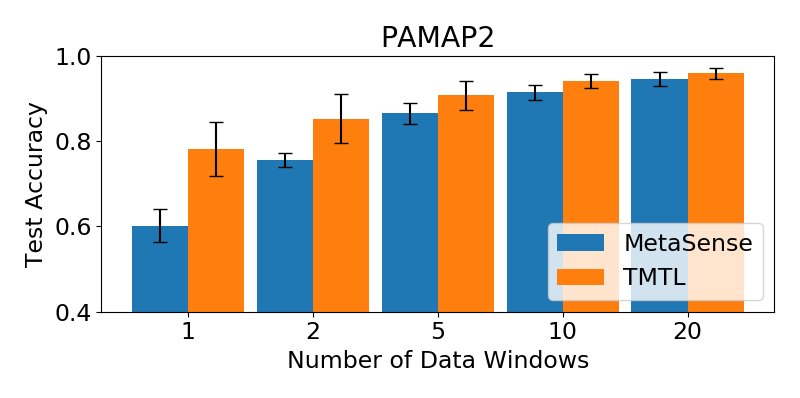}
			\includegraphics[width=0.49\linewidth]{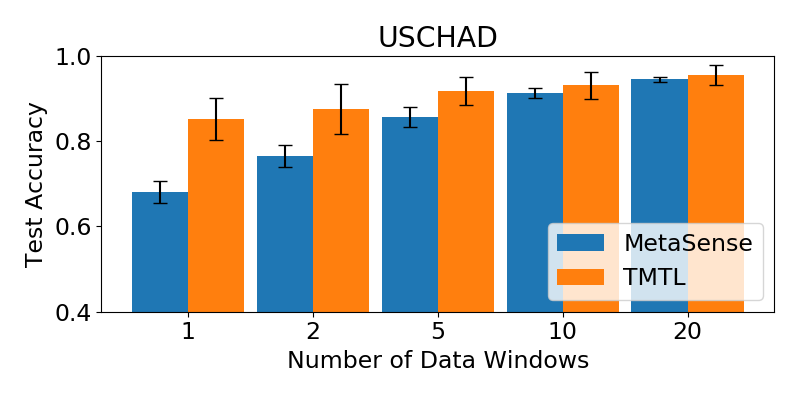}
			\includegraphics[width=0.49\linewidth]{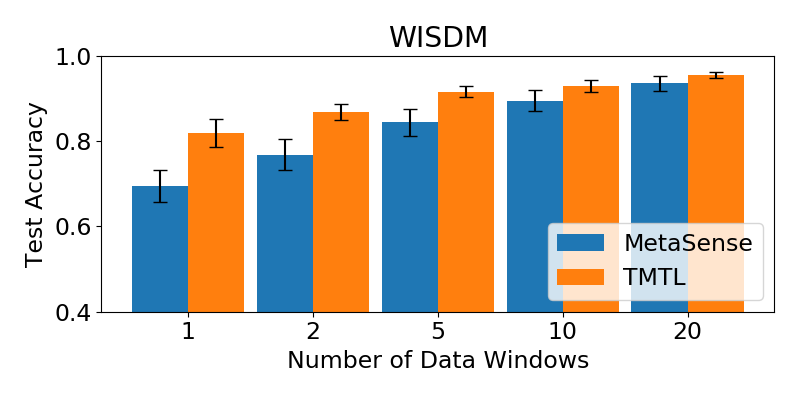}
			\includegraphics[width=0.49\linewidth]{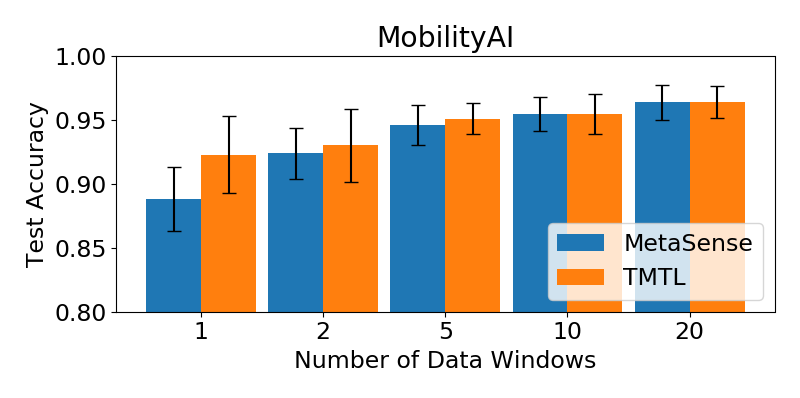}
		\end{minipage}
		\caption{\color{black}Comparison between TMTL and MetaSense on different domain adaptation tasks. Results are reported for 1, 2, 5, 10, 20-shots (data windows) per activity class from $\mathcal{D}_{tgt}$ with average test accuracy and standard error.}
		\label{fig:sota}
	\end{figure}\\
	
	\revtr{From Fig.\ref{fig:sota}, it is clear that in 19 out of 20 cases tested, TMTL performs better than or comparably as MetaSense. The advantage of TMTL is more prominent when few data samples are available from the target domain. The superior performance of TMTL over MetaSense for very few shot learning is due to the fact that MetaSense needs to update the entire model whereas TMTL only updates a task-specific layer. The latter approach is more data efficient and less likely to over-fit. However, as the number of available data samples from the target domain grows, the performance of MetaSense is comparable to or even slightly better than TMTL with sufficient labeled data from the target domain. 
		
		Compared to the TMTL with LSTM in Section 4.3.1, the performance of TMTL with LSTM drops by an average 2.35\% for 1-shot learning across all datasets. However, as the number of shots increases to more than 10, the performance gap is negligible ($<0.5\%$). This fact further demonstrates that IFLF is a model agnostic method and can be used in conjunction with any suitable network architecture for feature extraction.}
	\subsubsection{Similarity Metric and Fast Adaptation}
\begin{figure}[h!]
		\begin{minipage}[b]{\linewidth}
			\includegraphics[width=0.49\linewidth]{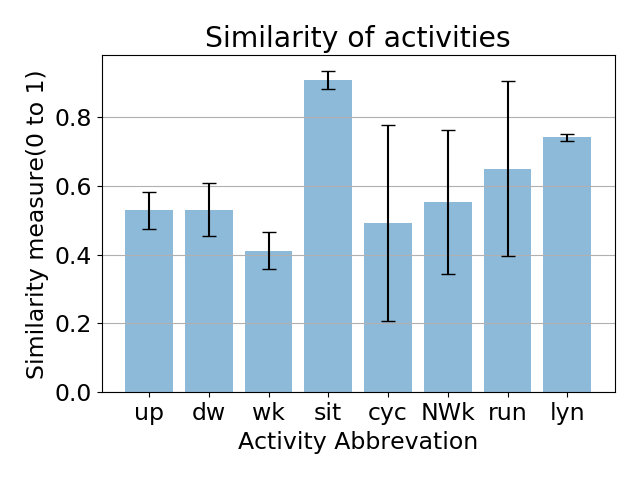}
			\includegraphics[width=0.49\linewidth]{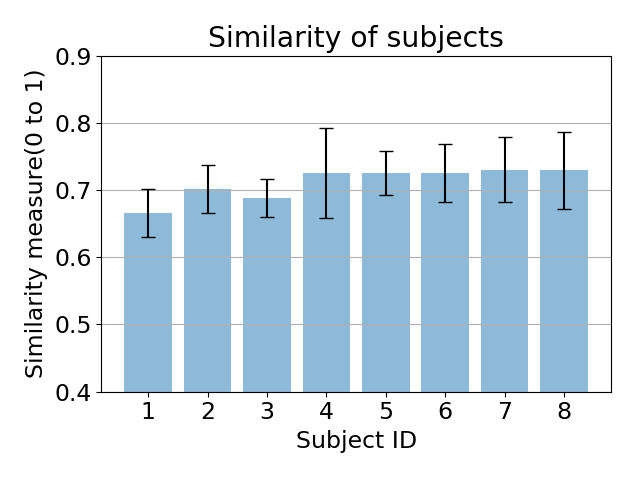}
			\subcaption{PAMAP2 dataset}
			\vspace{3mm}
			\includegraphics[width=0.49\linewidth]{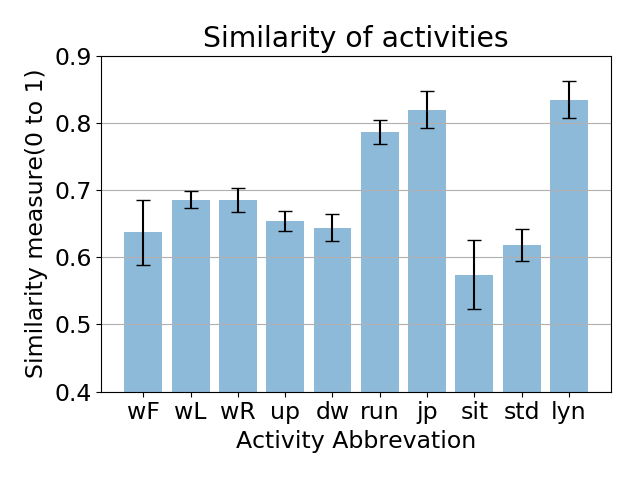}
			\includegraphics[width=0.49\linewidth]{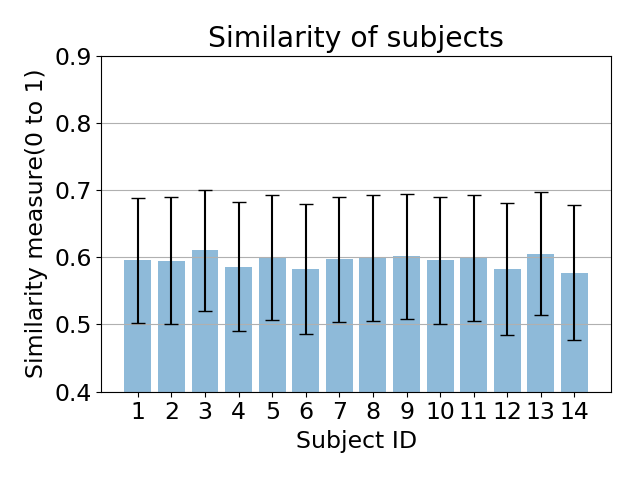}
			\subcaption{USCHAD dataset}
			\vspace{3mm}
			
			\includegraphics[width=0.49\linewidth]{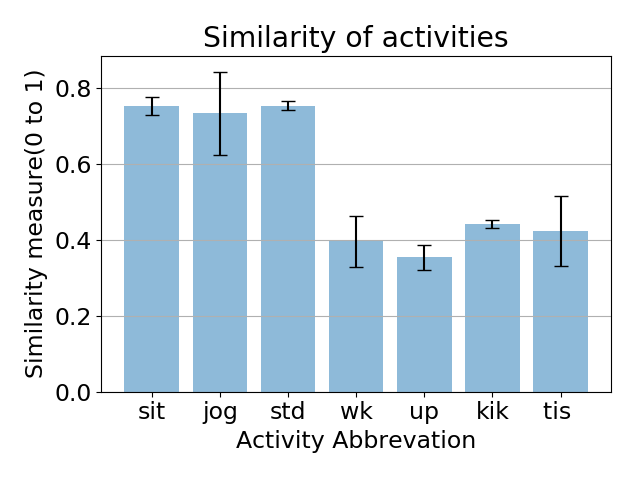}
			\includegraphics[width=0.49\linewidth]{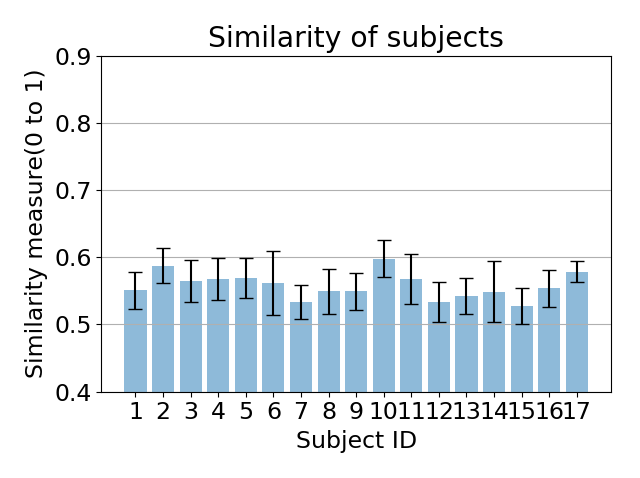}
			\subcaption{WISDM dataset}
			\vspace{3mm}
			
			\includegraphics[width=0.49\linewidth]{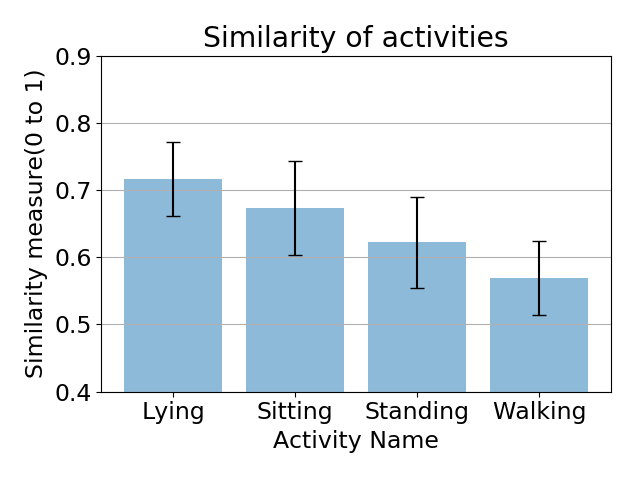}
			\includegraphics[width=0.49\linewidth]{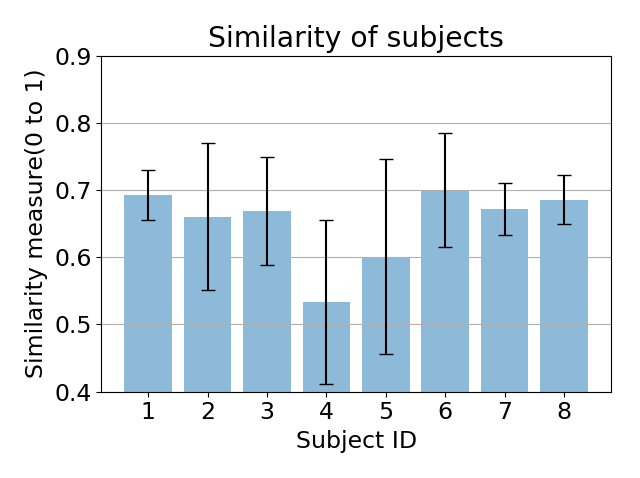}
			\subcaption{MobilityAI dataset}	
		\end{minipage}
		\caption{The evaluated similarity measure on each datasets. Plots on the left are similarity of activities, while plots on the right are similarity of subjects. Abbreviations in (a) (b) (c) are activities: \textit{upstairs, downstairs, walking, sitting, cycling, Nordic walking, running and lying}; \textit{walking forward, walking left, walking right, walking upstairs, walking downstairs, running forward, jumping up, sitting, standing and lying on bed}; \textit{sitting, jogging, standing, walking, upstairs, kicking soccer ball, playing tennis}}
		\label{fg:similarity_metric}
	\end{figure}

	Using the similarity metric defined in Section \ref{sec:3.4}, in this section, we first compare inter-subject and intra-activity similarity. The purpose of this study is to understand whether there exist subjects with similar movement patterns in all activities, and whether there exists an activity with little inter-subject variation. Secondly, we evaluate the use of intra-activity similarity in further improving data efficiency in fast adaptation. For activity similarity (left column in Fig. \ref{fg:similarity_metric}), we compute the averages and standard deviations of similarities of the same activity across different subjects. The right column in Fig. \ref{fg:similarity_metric} shows the similarity scores between one subject and the others averaged over matching activities.
	
	From Fig. \ref{fg:similarity_metric}, it is clear that some activities have higher similarity scores than the others. For example, in the PAMAP2 dataset, `sitting' appears to be similar across different subjects, while `walking' has the least similarity. This indicates diverse postures during walking but less variation during sitting in this dataset. On the other hand, when examining subject similarities, we find that some degree of similarity exists across almost all subjects (e.g., with a similarity score $> 0.5$). However, no two subjects perform activities the same way (e.g., the maximum similarity score is below 0.8). An exception is observed in the MobilityAI dataset, Subject 4 and 5 have noticeably lower similarity from others and larger standard deviations. It is because these two people have mobility issues and have to stand or walk with a rollator. It is expected that larger diversity exists among older adults with different underlying physical conditions. Homogeneity is more pronounced among younger populations as evident from the first three datasets. Another interesting observation is that depending on sensor placements, participants and the protocol of data collection, the same activity may have different cross-subject similarity in different datasets. For example, `lying on bed', `sitting' and `standing' have noticeably different similarity scores across the four datasets. 
	
\begin{figure}[h!]
		\begin{minipage}[b]{\linewidth}
			\includegraphics[width=0.48\linewidth]{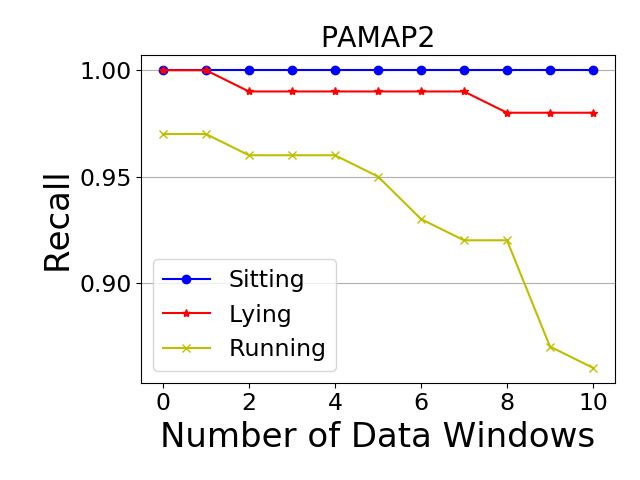}
			\includegraphics[width=0.48\linewidth]{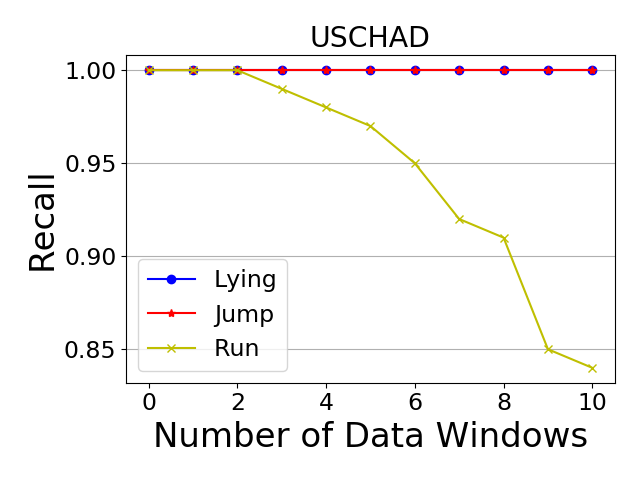}
			\includegraphics[width=0.48\linewidth]{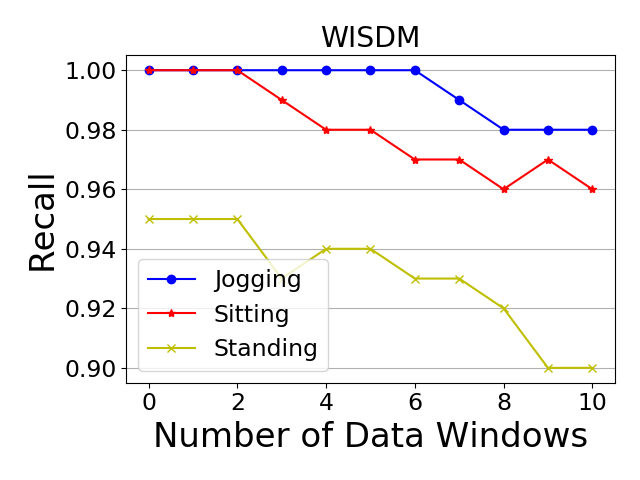}
			\includegraphics[width=0.48\linewidth]{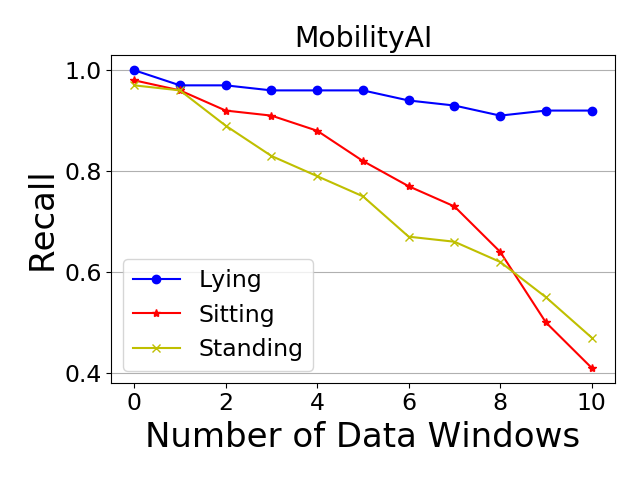}
		\end{minipage}
		\caption{The evaluation of applying similarity metric in the fast adaptation step. Activities with top 3 similarity scores are replaced in each dataset.}
		\label{fg:recall}
	\end{figure}

	Next, we investigate whether target domain data can be safely replaced by data from source domains for certain activities in fast adaptation. We first sample 10 labeled data windows from each activity from $\mathcal{D}_{tgt}$ to perform fast adaptation on a trained TMTL with the procedure described earlier. Then, the ten data windows of activities with top 3 similarity scores are replaced with data randomly sampled from $\mathcal{D}_{src}$ to update the TMTL model. The performance metric used is $recall=\dfrac{tp}{tp+fn}$, as it reveals whether a single activity is wrongly classified. A non-decreasing recall as more and more sampled data from the target domain is replaced by source domain data implies that doing so has little impact on the performance with less data collection efforts. The results for different datasets are shown in Fig. \ref{fg:recall}. 
	
	As evident from Fig. \ref{fg:recall}, `sitting' from PAMAP2, `lying' and `jump up' from USCHAD can be safely replaced with marginal impacts on the recall. This result is in accordance with the similarity scores presented in Fig. \ref{fg:similarity_metric} as all these three activities have similarity score $\geq0.8$. Although `lying' appears in three datasets, it can only be safely replaced in the USCHAD. In other words, by sampling activities with high similarity scores and small inter-subject variance from source domains, we can improve data efficiency by 12\% to 20\% on top of the reduction from fast adaptation.
	
	\section{Conclusion and future work}
	In this paper, we presented an invariant feature learning framework based on meta-training and multi-task learning paradigm to effectively address domain shifts and data shortage in HAR. \revtr{As demonstrated in Section \ref{sec:Results},} IFLF has been shown to work efficiently \revtr{in few-shot learning, especially when the number of shots are few (1 or 2 shots). A $>10\%$ performance margin has been observed when compared to MetaSense under such condition. Also, the proposed TMTL model implicitly handles class imbalance and class missing problems as well.} A similarity measure was proposed to further reduce the amount of data required in fast adaptation step. Though only domain shifts due to human and device variations have been considered in this work, we believe it can also be applied to handle sensor placement diversity, which will be investigated as part of our future work.
	
	Another research topic is to explore the application of domain generalization (DG) in sensor-based HAR. With the knowledge acquired from source domains, a model with DG can be directly utilized on the target domain without any data from it. \revtr{Replacing current relatively simple feature extractor with a more SOTA neural network such as DeepSense will be helpful to further improve the model generalization, and we will leave it as a future work.}
	
	\section*{Acknowledgment}
	We gratefully acknowledge the support of NSERC Discovery and CREATE programs with the funding for this research.

	\bibliographystyle{IEEEtran}
	\bibliography{IEEEabrv,ref}

\end{document}